# Two-parameter Leak Estimation in Non-invasive Ventilation

Francesco Vicario, *Member, IEEE*, Samiya Alkhairy, *Student Member, IEEE*,
Roberto Buizza, *Member, IEEE,* and William A. Truschel

*Abstract—* In this paper we present a method for the estimation of leaks in non-invasive ventilation. Accurate estimation of leaks is a key component of a ventilator, since it determines the ventilator performance in terms of patient-ventilator synchrony and air volume delivery. In particular, in non-invasive ventilation, the patient flow is significantly different from the flow measured at the ventilator outlet. This is mostly due to the vent orifice along the tube that is used for exhalation, but also to the non-intentional leaks that occur elsewhere in the circuit (e.g., at the mask). Such leaks are traditionally quantified via a model with two parameters, but only one of them is continually updated – the other is fixed. The new algorithm allows for breath-by-breath update of both parameters. This was made possible by leveraging a model describing the patient respiratory mechanics.

## I. INTRODUCTION

Non-invasive ventilation (NIV), also known as non-invasive positive pressure ventilation (NPPV), is widely used to assist breathing in both acute episodes of respiratory failure and chronic respiratory diseases [1], [2]. In contrast to invasive ventilation, NIV is characterized by a non-hermetic, open circuit design that is inherently leaky. These leaks pose a challenge in the respiratory therapy because they create a mismatch between the air flow supplied by the ventilator and the flow actually delivered to the patient. Accurate information about the patient's flow is crucial to i) to ensure that the patient receives sufficient tidal volume, ii) to guarantee that the lungs are not over distended (volutrauma) by excessive tidal volume, and iii) to improve the synchrony between the patient and the ventilator, since the ventilator typically detects the beginning of a patient breath from the estimated patient's flow. Usually, in NIV, leaks are both intentional and non-intentional. The former are necessary to allow for $CO_2$ removal and are created via an open orifice in the tube connecting the ventilator to the patient. The latter are due to imperfect adherence of the mask to the patient, mask misadjustments, mouth opening, etc. Intentional leaks can be taken into account, to some extent, by pre-characterization of the orifice. However, mucus secretions as well as water condensing in the circuit may alter the orifice characteristics over time. On the other hand, non-intentional leaks are hard to pre-characterize since they heavily depend on the individual patient and they dynamically change over time. For instance, non-intentional leaks tend to be minimal when the patient is awake. However, NIV is predominantly applied at night [3], when the loss of voluntary control and decreased muscle tone during sleep cause leaks to increase. Hence the need to dynamically estimate the leaks in order to accurately infer the patient's flow.

Due to the underdetermined nature of this mathematical problem, all methods for leak estimation in NIV need assumptions. For instance, [4] assumes that the mean patient's airflow over one or more breaths is zero. [5], [6] assume that leak flow is proportional to the square root of the pressure difference between the tube and the ambient. [5] also explicitly assumes that leakage does not vary quickly. [6], being based on a Kalman filter, implicitly makes assumptions on how quickly the amplitude factor of a leak model changes by specifying a value for its covariance. Regardless of their assumptions, existing methods are based on a model of the leaks of which only one parameter is updated over time. This choice is dictated by the fact that, given the nature of the problem, to estimate more parameters further assumptions are typically necessary.

We present a new algorithm for leak estimation whose main contribution is the continual update of two parameters of the leak model. At the core of the method is a physiological model of the patient's respiratory mechanics that is widely accepted in the medical community [7]. This model is the same as the one used in [6], but we take a step further and drastically reduce the number of assumptions concerning the patient. In [6] the resistance and compliance of the respiratory system had to be fixed, assumed a priori and never updated. Instead, we formulate a method that leverages the estimation scheme presented in [8] and, as a result, i) it does not require explicit knowledge of the resistance and compliance, ii) it features the respiratory time constant as the only physiological parameter of the patient's lung mechanics needed for leak estimation, iii) it estimates such parameter on a breath-by-breath basis, overcoming the limitation of using fixed values, often derived from population averages that poorly represent the specific individual. Finally, we validate the concept using experimental data from a lung simulator where the patient's flow (or the leaks) can be directly measured.

## II. METHOD

With reference to the simplified schematics in Fig. 1, where the leaks are lumped into a single term ($Q_l$), we want to estimate the flow ($Q_p$) inhaled and exhaled by the patient i)

F. Vicario (corresponding author - phone: +1-617-613-2262; fax: +1-914-945-6580; e-mail: francesco.vicario@philips.com) and R. Buizza (roberto.buizza@philips.com), are with Philips Research North America, Cambridge, MA 02114.

S. Alkhairy (samiya@mit.edu) is with the Massachusetts Institute of Technology, Cambridge, MA 02139. At the time this work was done, she was with Philips Research North America.

W. A. Truschel (bill.truschel@philips.com) is with Philips Respironics, Monroeville, PA 5146.

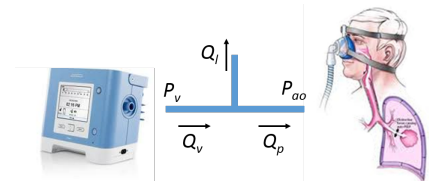

Figure 1. Schematic representation of the system under consideration.

from pressure ($P_v$) and flow ($Q_v$) waveforms measured at the ventilator outlet, ii) in the presence of leaks ($Q_l$) in the circuit between the ventilator and the patient, and iii) breath-by-breath (real-time).

## A. Model of Leak Flow

The leaks are typically modeled as

$$Q_l(t) = G_{orf} P^\gamma(t) \quad (1)$$

where $G_{orf}$ and $\gamma$ are two parameters assumed to be constant over one breath and $P$ is a pressure representative of the pressure at which the leaks occur along the circuit. The estimate of $P$ can be fine-tuned if the tubing has been characterized, but in this work, for simplicity, we will assume $P = P_v$. The continuity equation along the circuit is

$$Q_p(t) = Q_v(t) - Q_l(t) \quad (2)$$

From (2) it is evident how, to estimate $Q_p$, we have to estimate $Q_l$ or, in turn, $G_{orf}$ and $\gamma$. Integrating (2) over a breath of duration $T$ and assuming that the volume of air (integral of the flow) inhaled by the patient equals the one exhaled, we can write

$$0 = \int_0^T Q_v(t)dt - \int_0^T Q_l(t)dt \quad (3)$$

Plugging (1) into (3) and solving for $G_{orf}$, we obtain

$$G_{orf} = \int_0^T Q_v(t)dt \, / \int_0^T P_v^\gamma(t)dt \quad (4)$$

which is the classic equation to estimate leaks in NIV (see, e.g., [6]). However, it requires one to assume a value for $\gamma$. As a consequence, current methods update breath-by-breath only $G_{orf}$ while leaving $\gamma$ fixed (typically equal to 0.5 according to the assumption of turbulent flow). This is suboptimal since the assumption that the flow is turbulent and that the leak is a fixed geometry can lead to errors in estimation with many examples of common NIV interfaces. In this paper, we want to develop an algorithm that updates breath-by-breath the estimate of leaks by updating both $G_{orf}$ and $\gamma$. Besides (4), another equation is necessary in order to construct a set of two equations in two unknowns.

## B. Model of Respiratory Mechanics

We exploit the first-order single-compartment linear model [7] of the respiratory system, whose electrical analogue is shown in Fig. 2a. The model is characterized by two parameters, $R$ and $E$, that represent the respiratory resistance, mainly due to the airways, and elastance, mainly due to the lungs and the chest wall. $P_{ao}$ is the pressure provided by the ventilator at the patient's airway opening (typically, the mouth). The patient's respiratory drive is account for via an equivalent pressure $P_{mus}$. The overall dynamics of the patient's air flow $Q_p$ is governed by the so-called equation of motion of the respiratory system

$$P_{ao}(t) = RQ_p(t) + E\int_0^t Q_p(s)ds + P_{mus}(t) + P_0 \quad (5)$$

where the integral corresponds to the volume of air entering the patient's lungs from the start of the breath ($t = 0$), and $P_0$ is a constant pressure term balancing the pressure provided by the ventilator at $t = 0$. Like for (1), we assume for simplicity that $P_{ao} = P_v$. For each breath, consider only the portion of the exhalation between $t_i$ and $t_f$ where these two time samples are the first and last times during the breath, respectively, where the ventilator maintains the set positive-end expiratory pressure, so that $P_v(t) = PEEP$ for all $t_i \leq t \leq t_f$. During exhalation we can assume $P_{mus} = 0$ even for spontaneously breathing patients. Plugging (1) into (2), (2) into (5) and rearranging terms conveniently, we obtain

$$\int_{t_i}^t Q_v(s)ds = -\tau Q_v(t) + Q_z t + K \quad \text{for } t_i \leq t \leq t_f \quad (6)$$

where $\tau$, $Q_z$, and $K$ are constants. $\tau = R/E$ is the respiratory time constant, $Q_z = G_{orf} PEEP^\gamma$ is $Q_l(t)$ for $t_i \leq t \leq t_f$, and $K$ incorporates all the constant terms arising in the derivation of (6). Writing (6) for all time samples $t_i \leq t \leq t_f$, the following set of equations in matrix form is obtained

$$\begin{bmatrix} -Q_v(t_i) & t_i & 1 \\ \vdots & \vdots & \vdots \\ -Q_v(t_f) & t_f & 1 \end{bmatrix} \begin{bmatrix} \tau \\ Q_z \\ K \end{bmatrix} = \begin{bmatrix} \int_{t_i}^{t_i} Q_v(s)ds \\ \vdots \\ \int_{t_i}^{t_f} Q_v(s)ds \end{bmatrix} \quad (7)$$

The three unknown constants in (7) can be estimated by the ordinary least-squares (LS) method. The estimation of one of them, $Q_z$, provides the necessary relationship to solve for both parameters in (1).

## C. Estimation Algorithm

The leak estimation algorithm then requires to i) fit (6) to the measured data to obtain an estimate of $Q_z$ (and $\tau$, if desired), which can be done by computing the pseudo-inverse of the data matrix in (7), ii) solve numerically the following set of two equations for $G_{orf}$ and $\gamma$

$$G_{orf} \int_0^T P_v^\gamma(t)dt = \int_0^T Q_v(t)dt \quad (8a)$$

$$Q_z = G_{orf} PEEP^\gamma \quad (8b)$$

and, iii) estimate $Q_l$ and $Q_p$ from (1) and (2), respectively. The key idea of obtaining preliminary information by fitting the exhalation data was inspired by the respiratory mechanics estimation algorithm in [8].

## C. Lung Simulator

Like in [8], the lung simulator in Fig. 2b is used to test the new estimation method. The lung simulator includes an elastic balloon (Model 5432, Hudson RCI) representative of the lung and chest wall. The balloon is connected to a

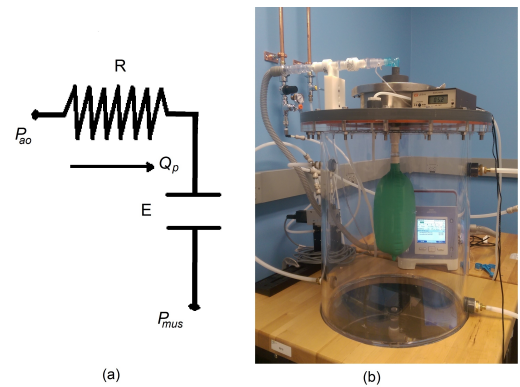

Figure 2. Electrical analogue of the respiratory mechanics model (a) and lung simulator used to test the leak estimation algorithm (b) [8].

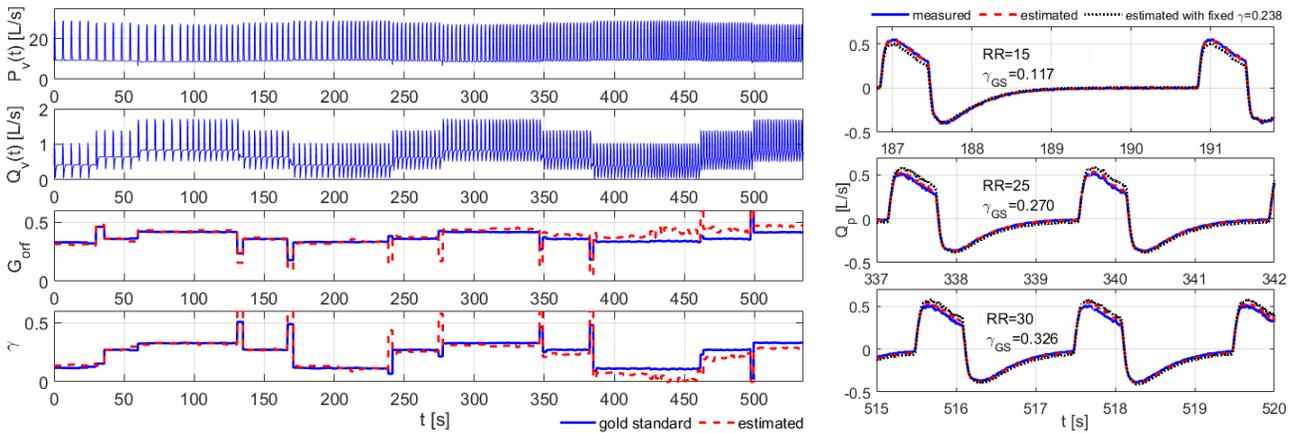
Figure 3. Dataset 1: changes in respiratory rate with plateau exhalation valve (left) and examples of $Q_p$ waveforms (right).

Trilogy ventilator (Philips Respironics) by a tube to which a pneumatic resistor is added. Two different resistors are available, with nominal resistances of 5 and 20 cmH$_2$O·s/L (Model 7100R-5 and -20, Hans Rudolph). The balloon is enclosed in a chamber whose pressure is controlled to be $P_{mus}$ by a valve (Model MPYE-5-1/4-010-B, Festo) connected to compressed air and vacuum. Two different exhalation orifices are used: a whisper swivel and a plateau exhalation valve, both from Philips Respironics. $Q_v$ and $Q_p$ are measured via flow sensors (Pneumotach Amplifier 1 series 1110, Hans Rudolph) located right before and after the vent orifice. $P_v$ is measured near the orifice. The data from all the sensors are collected real-time via an xPC Target™ connected to a laptop. The leak estimation algorithm runs within Matlab®. Measuring $Q_p$ provides us with a ground truth waveform and derived parameters against which the estimates of the algorithm are validated.

## III. RESULTS

Three datasets are shown to demonstrate the effectiveness of the proposed technique. The tests are performed using the Trilogy ventilator in T mode, i.e., delivering pressure-controlled mandatory breaths. All datasets are generated with the ventilator set to deliver breaths with inhalation time of 0.6 s. The inspiratory pressure and PEEP are set to 30 and 10 cmH$_2$O, respectively, for the first dataset and 25 and 10 for the other two. Across each dataset, the respiratory rate is varied from 10 to 30 bpm (breaths per minute).

The first dataset (Fig. 3) is obtained with the plateau exhalation valve. This type of orifice is ideal to test our algorithm because it features two additional lateral ports that can be used to change the orifice values of both $\gamma$ and $G_{orf}$. At each respiratory rate value, a few breaths are collected in each configuration (both ports closed, one port open, both ports open). In the ventilator flow plot in Fig. 3, opening (closing) a port causes a shift upwards (downwards) of the waveforms. The gold standard (GS) values of $G_{orf}$ and $\gamma$ in Fig. 3 are computed by fitting (1) to the data via the LS method, with $Q_l$ computed as the difference between the measured flows $Q_v$ and $Q_p$. The estimates of both $G_{orf}$ and $\gamma$ obtained by the algorithm presented in this paper track the changes in the corresponding gold standard values, proving the capability of updating both parameters on-line and thus improving the estimation of the flow actually delivered to the patient. Note that the spikes at the breaths corresponding to sudden changes in $G_{orf}$ and $\gamma$ in Fig. 3 should be ignored, since they are artifacts due to the manipulation of the orifice ports. Across the dataset, it is evident how the accuracy of the estimates of $G_{orf}$ and $\gamma$ decreases as the respiratory rate increases. An examination of the plots to the right may provide further insight. They show examples of estimated patient flow waveforms from three different regions of respiratory rate (from top to bottom, 15, 25, and 30 bpm). Since the inhalation time is maintained constant, higher respiratory rate reduces the exhalation time and $Q_p$ does not have the time to show the asymptote at 0. Correspondingly, with decreased exhalation time the waveform of $Q_v$ does not show its complete exponential profile, making it more challenging for the first step of the algorithm to provide an accurate estimate of $Q_z$ and $\tau$. Despite the lower estimation accuracy for $G_{orf}$ and $\gamma$ at high respiratory rates, the proposed algorithm provides better $Q_p$ estimates than a traditional algorithm based on (4), that estimates only $G_{orf}$ and relies on a pre-determined, fixed value of $\gamma$. In Fig. 3 (right plots), the waveforms that would be estimated using a fixed value $\gamma = 0.238$, corresponding to the average gold standard $\gamma$ across the entire dataset, are shown to be less accurate.

For further validation, the second dataset is acquired using the whisper swivel as orifice (Fig. 4), which features different $G_{orf}$ and $\gamma$ than the orifice used in Fig. 3. Similar to what observed above, the parameters of the leak model are accurately estimated, except for when the exhalation duration is significantly reduced. Finally, to show how the new method is capable of adapting to patients with different respiratory mechanics parameters, we show in Fig. 5 the results obtained performing the same experiment as in Fig. 4 but changing the pneumatic resistor in the test bench from 20 to 5 cmH$_2$O·s/L. Note how the decreased resistance makes the patient's exhalation faster (smaller respiratory time constant), to the benefit of the estimation accuracy even at higher respiratory rates.

## IV. DISCUSSION

In this paper we presented a method to enhance the estimation of the leaks between the ventilator and the lungs and, in turn, of the patient flow. The estimation of the net flow that reaches the patient's respiratory system is crucial to guarantee both adequate and safe ventilation. Accurate patient flow estimation will also improve the synchrony between the patient and the ventilator, since the ventilator typically detects the beginning of a patient's breath from this estimation. This is particularly important for open-valve (or valve-less) ventilators, which is typically the case in NIV. However, it is worth noting that the importance of accurate leak estimation goes beyond NIV. Although leaks are more significant in NIV due to the presence of an exhalation orifice and imperfect interfaces (oral, nose or full masks), leaks are often present in invasive ventilation, too. For instance, intubated patients may experience leaks around the cuffs of the endotracheal or tracheostomy tube.

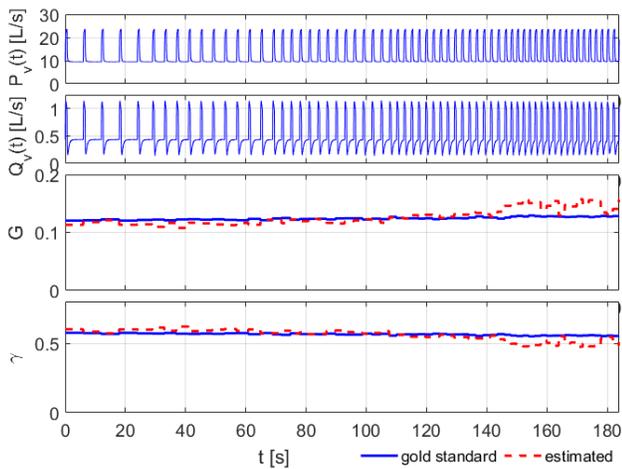

Figure 4. Dataset 2: changes in respiratory rate with whisper swivel.

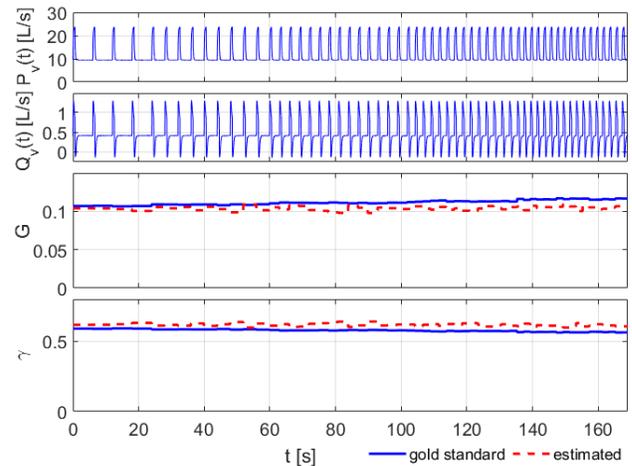

Figure 5. Dataset 3: same as dataset 2 except for decreased resistance.

The method presented in this paper aims to improve leak estimation by updating breath-by-breath both parameters in the leak model. The examples demonstrate that the continual and simultaneous update of both parameters is indeed possible and, as expected, improves the estimates over methods that update only one parameter (Fig. 3, right plots). At the core of the new method is a physiological model of the patient's respiratory mechanics. As opposed to methods that need to assume the values of the parameters of such a model and keep them fixed, this new technique is designed to naturally adapt to different patients, as simulated in the test bench experiments with a change of respiratory resistance (Figs. 4 and 5). Therefore, the method is suitable for patients whose parameters change over time, either quickly due to occasional events or slowly because of, e.g., the progression of the disease.

Among the limitations of the method, we found that with short exhalation times the flow waveform may not carry sufficient information and this may affect the algorithm performance. Additionally, the method might need to be modified for patients with diseases that significantly alter the respiratory mechanics making it not adequately represented by the linear model in Fig. 1a. In choosing different models, the authors recommend using lumped-parameter models, like the one used in this paper, as opposed to models with a large number of parameters to be estimated, like finite-element models. For instance, a model with collapsible airways like the one in [9] can be considered for COPD patients suffering from expiratory flow limitation.

While the method was investigated in detail using a test bench representative of the human lung mechanics, we anticipate further work to assess the benefit of this enhancement of leak estimation with actual patients. We must note that the method's performance in human subjects will be assessed using a different set of metrics that does not require using direct measurement of the leaks (or, like in the examples in the previous section, of the patient's flow). While these direct flow measurements are achievable on a test bench where the leaks were concentrated in a single orifice and the rest of the system was tested to be airtight, this is not the case with patients, where leaks can occur at multiple locations in the circuit between the ventilator and the patient (e.g., mask or endotracheal tube), and hence it is difficult to make direct measurements of leaks with the same accuracy and reliability. To overcome this technical problem, future work will attempt to quantify the benefit in human subjects, for instance, in terms of improved patient-ventilator synchrony (like in [6]), improved oxygenation and $CO_2$ removal, and reduced ventilator-induced lung injuries, which, in turn, give better patient's outcomes.